\documentclass[aps,pre,10pt,showpacs,floatfix,onecolumn,nofootinbib,a4paper]{revtex4-2}
\usepackage{amssymb, amsmath, amsthm, mathtools}
\usepackage{bm}
\usepackage{mathrsfs}
\usepackage{hyperref,bookmark}
\usepackage{graphicx}
\usepackage{epsfig,color}
\usepackage{float}
\usepackage{subcaption}
\usepackage{verbatim}
\usepackage{dcolumn}
\usepackage{tikz}
\usepackage{url}

\usepackage[utf8x]{inputenc}
\usepackage[english]{babel}
\usepackage{mathrsfs}
\usepackage{epsfig}
\usepackage{listings}

\usepackage{mdwlist}
\usepackage{bm}
\usepackage{dsfont}
\usepackage{oldgerm}
\usepackage{geometry}
\usepackage{relsize}
\usepackage{afterpage}
\usepackage{lmodern}
\usepackage{tabularx}
\usepackage{cancel}

\newcommand{\dd}{\mathrm{d}}

\newcommand{\vt}{\vartheta}
\newcommand{\vp}{\varphi}
\newcommand{\trans}{^\mathsf{T}}

\newcommand{\n}{\bm{l}_1}
\newcommand{\m}{\bm{r}_1}
\newcommand{\nper}{\bm{l}_2}
\newcommand{\mper}{\bm{r}_2}

\newcommand{\e}{\bm{e}}
\newcommand{\x}{\bm{x}}
\newcommand{\zero}{\bm{0}}
\newcommand{\y}{\bm{y}}

\newcommand{\normal}{\bm{\nu}}
\newcommand{\surface}{\mathscr{S}}

\newcommand{\nablas}{\nabla\!_\mathrm{s}}

\newcommand{\uv}{\bm{u}}

\newcommand{\vv}{\bm{v}}
\newcommand{\W}{\mathbf{W}}

\newcommand{\I}{\mathbf{I}}

\newcommand{\R}{\mathbf{R}}
\newcommand{\C}{\mathbf{C}}
\newcommand{\B}{\mathbf{B}}
\newcommand{\U}{\mathbf{U}}
\newcommand{\V}{\mathbf{V}}
\newcommand{\Lt}{\mathbf{L}}
\newcommand{\M}{\mathbf{M}}

\newcommand{\tr}{\operatorname{tr}}
\newcommand{\co}{\operatorname{co}}

\newcommand{\nay}{(\nabla\y)}
\newcommand{\naya}{(\nabla\y^\ast)}
\newcommand{\nan}{(\nabla\normal)}
\newcommand{\nana}{(\nabla\normal^\ast)}
\newcommand{\curvature}{(\nablas\normal)}
\newcommand{\curvaturast}{(\nablas\normal^\ast)}
\newcommand{\A}{\mathbf{A}}

\newcommand{\euclid}{\mathscr{E}}
\newcommand{\framec}{(\e_1,\e_2,\e_3)}

\newcommand{\tangent}{\bm{t}}

\newcommand{\Rnu}{\mathbf{R}_{\normal}}
\newcommand{\pure}{\mathbb{A}}
\newcommand{\orth}{\mathsf{SO}(3)}
\newcommand{\orthnu}{\mathsf{SO}(\normal)}
\newcommand{\proj}{\mathbf{P}(\normal)}
\newcommand{\Upm}{(\mathbf{U}')^{-1}}

\newcommand{\nigh}[1]{{\color{black}{#1}}}

\begin{document}
	\title{Pure measure of  bending for soft plates}
	\author{Epifanio G. Virga}
	\email{eg.virga@unipv.it}
	\affiliation{Dipartimento di Matematica, Universit\`a di Pavia, Via Ferrata 5, 27100 Pavia, Italy }

	\date{\today}

	\begin{abstract}
	This paper, originally motivated by a question raised by Wood and Hanna [\emph{Soft Matter}, \textbf{15}, 2411 (2019)], shows that pure measures of bending for soft plates can be defined by introducing the class of \emph{bending-neutral} deformations, finite incremental changes of the plate's shape bearing \emph{no} further bending. This class of deformations is subject to a geometric compatibility condition, which is fully characterized. A tensorial pure  measure of bending, which is accordingly invariant under bending-neutral deformations, is described in details. \nigh{As shown by an illustrative class of examples,} the general notion of pure measure of bending could be of use to formulate direct theories for soft plates, where stretching and bending energies are kept separate.
	\end{abstract}
	
	\maketitle

\section{Introduction}\label{sec:intro}
It has been remarked that a \emph{pure} measure of bending for extensible rods \emph{cannot} be provided by the curvature of its centerline, as this can easily been shown to be affected by a superimposed dilation \cite{wood:contrasting}. Actually, within the nonlinear theory of extensible  elastic rods, this issue has long be known, at least since the work of Antman \cite{antman:general} and others who soon followed in his footsteps \cite{reissner:one-dimensional,whitman:exact}. It has also consequences on the choice of the appropriate form of the bending content of the strain energy, when in the direct approach one wishes that it be  nicely separated from the stretching content.

This was indeed the criterion  advocated in \cite{antman:general} for a planar extensible \emph{elastica}: if $\vt$ denotes the centerline's deflection angle (relative to a fixed direction in the plane) of a naturally straight rod, the appropriate quadratic bending energy independent of stretching would be proportional to $(\partial_x\vt)^2$, and not $(\partial_s\vt)^2$, where $x$ and $s$ denote the arc-length coordinate in the reference and present configurations, respectively. As also remarked in \cite{antman:general}, a theory measuring the bending energy of an extensible rod by $(\partial_s\vt)^2$, such as that proposed in \cite{tadjbakhsh:variational}, would in general be more complicated.\footnote{It seems indeed simplicity the major advantage of Antman's proposed constitutive law, as witnessed by the neat analysis that ensued in \cite{antman:general}.}

Other attempts to derive the balance equations for an extensible elastic rod have been based in \cite{magnusson:behaviour} and \cite{oshri:strain} on Biot's \emph{nominal} strain tensor \cite{biot:non-linear} (see also \cite{truesdell:mechanical} and \cite{biot:mechanics}); they corroborated the approach purported in \cite{antman:general}. A similar corroboration came from \cite{irschik:continuum}, which followed yet another avenue, designed according to the canons of modern continuum mechanics.

Similarly, for a plate whose planar midsurface $S$ in the reference configuration  gets deformed into the surface $\surface$ embedded in three-dimensional space, the invariants of the curvature tensor $\nablas\normal$, where $\normal$ is the outer unit normal to $\surface$ and $\nablas$ denotes the surface gradient, cannot be pure measures of bending, as also neatly illustrated in \cite{vitral:dilation}.

In analogy with the theory of stretchable rods, proposals for the bending energy of plates and shells (i.e., plates curved in their natural state) have been advanced that are not affected by one or another form of stretching. The lines of thought followed to achieve this goal are disparate, as are the conclusions reached; controversies abound, especially for shells, which I deliberately leave out of the scope of this paper.

In an alternative approach to geometric elasticity as formulated in \cite{efrati:elastic}, a theory for deformable, geometrically incompatible sheets was derived in \cite{oshri:strain} from a three-dimensional model phrased in terms of Biot's strain measures. The outcome is in contrast with \cite{efrati:elastic}, also in the rod-limit case, where the alternative energy is closer to Antman's. The theory of \cite{oshri:strain} has been applied to a soft matter system in \cite{oshri:modeling}; it effectively employs a measure of bending for plates that has a distant antecedent in \cite{atluri:alternate}.  

More recently, a dimension reduction method following the general lines proposed in \cite{steigmann:thin,steigmann:two-dimensional,steigmann:extension,steigmann:koiter} was applied  in \cite{vitral:energies} to a three-dimensional isotropic quadratic strain energy \cite{vitral:quadratic} formulated in terms of Biot's strain tensor  to justify a model for shells. In particular, in the limiting case of plates, the one that concerns us here, two two-dimensional bending measures arose naturally from the method employed in \cite{vitral:energies}, one invariant under arbitrary superimposed stretchings, and the other invariant only under superimposed dilations, both reducing to Antman's in the rod-limit.

The questions addressed in this paper will be of a pure kinematic nature. First, we need to identify  a criterion of \emph{bending-neutrality} for soft (highly deformable) plates: we must single out the incremental deformations that, although changing the shape of $\surface$, do so without any further bending. Thus, \emph{pure} measures of bending will be defined as the ones \emph{invariant} under bending-neutral deformations. I shall focus on one isotropic such measure, $\A$, tensorial in character, alongside with its scalar invariants and contrast them with both similar and dissimilar measures from the literature. It will be shown that $\A$ reduces to Antman's measure for extensible rods. 

Identifying the pure measure of bending appropriate to a class of materials  should prelude to  a proposal for a direct theory of plates for which stretching and bending modes are kept separate from one another.

This paper is organized as follows. In Sec.~\ref{sec:pure}, we define the class of bending-neutral deformations and present an isotropic measure of pure bending associated with them. Section~\ref{sec:compatibility} is devoted to the compatibility condition that a bending-neutral deformation must obey; we shall see that this class contains both pure tangential stretching deformations and planar \emph{drill} rotations, characterized separately in \cite{szwabowicz:pure} and \cite{saem:in-plane}. Finally, in Sec.~\ref{sec:conclusions}, we summarize the conclusions of this study and comment briefly about its possible extensions.

\section{Bending-Neutral Deformations}\label{sec:pure}
\begin{figure}[h]
	\begin{center}
		\includegraphics[width=.6\linewidth]{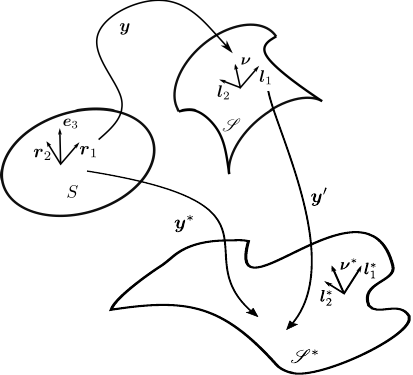}
	\end{center}
	\caption{\label{fig:sketch} A flat surface $S$ in the $(x_1,x_2)$  plane of a fixed Cartesian frame $\framec$ is deformed by the mapping $\y$ into a smooth surface $\surface$ embedded in three-dimensional Euclidean space $\euclid$. The unit vectors $(\m,\mper)$ are the right principal directions of stretching in the reference configuration, while $(\n,\nper)$ are the left principal directions of stretching  in the present configuration; $\e_3$ is the outer unit normal to $S$, while $\normal$ is the outer unit normal to $\surface$;  they are oriented so that $\e_3=\m\times\mper$ and $\normal=\n\times\nper$, respectively. Surface $\surface^\ast$ is obtained by combining $\y$ with the incremental deformation $\y'$, so that $\y^\ast=\y'\circ\y$. The outer unit normal to $\surface^\ast$ is $\normal^\ast$ and the left principal directions of stretching associated with $\y^\ast$ are $\n^\ast$ and $\nper^\ast$, with $\normal^\ast=\n^\ast\times\nper^\ast$.}
\end{figure}

Consider a surface $S$ lying on the $(x_1,x_2)$ plane of a Cartesian frame $\framec$, so that $\e_3$ is one of its unit normals. Let $\x$ denote the position vector in $S$ and let $\y$ be the deformation that maps $S$ onto $\surface$ in three-dimensional Euclidean space $\euclid$. We shall assume that $\y$ is at least of class $C^2$ and we shall denote by $\normal$ the unit normal to $\surface$ oriented coherently with the orientation of $\e_3$ (see Fig.~\ref{fig:sketch}).

Extending to the present geometric setting results  well-known from three-dimensional kinematics (see, e.g., \cite[Ch.\,6]{gurtin:mechanics}), the deformation gradient $\nabla\y$ can be represented as
\begin{equation}\label{eq:representation_nabla_y}
\nabla\y=\lambda_1\n\otimes\m+\lambda_2\nper\otimes\mper,
\end{equation}
where $\nabla$ denotes the gradient in $\x$, the positive scalars $\lambda_1$, $\lambda_2$ are the principal stretches and the unit vectors $\bm{r}_i$, $\bm{l}_i$ are the corresponding \emph{right} and \emph{left} principal directions of stretching. While $\m(\x)$ and $\mper(\x)$ live on the $(x_1,x_2)$ plane for all $\x\in S$, $\n(\x)$ and $\nper(\x)$ live on the tangent plane to $\surface$ at $\y(\x)$. The right and left Cauchy-Green tensors are correspondingly defined and represented by 
\begin{subequations}\label{eq:representations}
	\begin{align}
		\C:=\nay\trans\nay&=\lambda_1^2\m\otimes\m+\lambda_2^2\mper\otimes\mper,\label{eq:representation_C}\\
		\B:=\nay\nay\trans&=\lambda_1^2\n\otimes\n+\lambda_2^2\nper\otimes\nper.\label{eq:representation_B}
	\end{align}	
\end{subequations}

I wish to introduce a \emph{bending-neutral} deformation as an incremental (finite) deformation $\y'$ that maps $\surface$ into a surface $\surface^\ast$ \emph{without} further bending it. To make this notion precise, we first
consider the composed deformation $\y^\ast:=\y'\circ\y$, where $\y'$ is a general, smooth (at least $C^2$) mapping that changes $\surface$ into $\surface^\ast$ and is characterized by having 
\begin{equation}
	\label{eq:nabla_y'}
	\nabla\y'=\R'\U',
\end{equation}
where $\R'$ belongs to the proper orthogonal group in three dimensions, $\orth$, and
\begin{equation}
	\label{eq:U'}
	\U'=\lambda_1'\uv'_1\otimes\uv'_1+\lambda_2'\uv'_2\otimes\uv'_2
\end{equation}
is a  \emph{stretching} tensor with principal directions of stretching $\uv'_1$, $\uv'_2$ tangent to $\surface$ at $\y(\x)$ and principal stretches $\lambda_1'$, $\lambda_2'>0$. The tensor $\U'$ is to be regarded as a (linear) mapping of the tangent plane to $\surface$ at $\y(\x)$ into itself.

Equation \eqref{eq:nabla_y'} is the form of polar decomposition fit to describe the deformation of material surfaces; it was established in \cite{man:coordinate} within the general coordinate-free theory introduced in \cite{gurtin:continuum}, which will also be followed here (see also \cite{pietraszkiewicz:determination}). A tensor $\V'$ that maps the tangent plane to $\surface^\ast$ at $\y^\ast(\x)=\y'(\y(\x))$ into itself, can  also be introduced via the following equation,
\begin{equation}
	\label{eq:polar_U_V}
	\R'\U'=\V'\R'.
\end{equation}
It follows from \eqref{eq:polar_U_V} that $\V'$ possesses the same eigenvalues $\lambda_1'$, $\lambda_2'$ as $\U'$, while its eigenvectors are  $\R'\uv'_1$, $\R'\uv'_2$, which generally differ from the left principal directions of stretching $\n^\ast$, $\nper^\ast$ associated with $\y^\ast$  (and shown Fig.~\ref{fig:sketch}). By the chain rule,
\begin{equation}
	\label{eq:chain_rule}
	\nabla\y^\ast(\x)=\nabla\y'(\y(\x))\nabla\y(\x)\quad\forall\ \x\in S,
\end{equation}
\eqref{eq:nabla_y'} and \eqref{eq:polar_U_V} imply that
\begin{equation}
	\label{eq:B_ast}
	\B^\ast:=\naya\naya\trans=\V'(\R'\B\R^{\prime\mathsf{T}})\V',
\end{equation}
which shows that the eigenframes of $\B^\ast$ and $\V'$ coincide whenever $\B$ and $\U'$ commute.\footnote{This follows easily from applying both sides of \eqref{eq:B_ast} to the eigenvectors $\R'\uv'_i$ of $\V'$.}

It should also be noted that since $\y'$ is defined on $\surface$, its gradient $\nabla\y'$ featuring in \eqref{eq:nabla_y'} is to be properly regarded as a \emph{surface} gradient on $\surface$, $\nablas\y'$, according to the definition given in \cite{gurtin:continuum} (see also \cite{murdoch:coordinate}). 

We say that $\y'$ is a \emph{bending-neutral} deformation if
\begin{equation}
	\label{eq:bending-neutrality_definition}
	\R'=\R_0\Rnu,
\end{equation}
where $\R_0\in\orth$ is a \emph{uniform} rotation, i.e., independent of position, and $\Rnu\in\orthnu$, where $\orthnu$ is the group of rotations about the unit normal $\normal$. While $\R_0$ is a global \emph{rigid motion}, $\Rnu$ is a local \emph{drill} rotation. For $\R'$ as in \eqref{eq:bending-neutrality_definition}, \eqref{eq:nabla_y'} represents an incremental deformation that entails local \emph{stretching} of $\surface$, described by $\U'$, possibly followed by a \emph{twist} distortion about $\normal$, described by $\Rnu$, and an overall undistorting rotation $\R_0$.

We shall discuss in Sec.~\ref{sec:compatibility} the compatibility condition that a bending-neutral deformation $\y'$ must obey to exist, at least locally, on $\surface$. Here we only heed that this class of incremental deformations includes both pure stretching deformations, for which $\Rnu$ is the identity, and pure drill rotations, for which $\U'$ is the projection $\proj:=\I-\normal\otimes\normal$ on the plane orthogonal to $\normal$.\footnote{$\I$ is the identity tensor in three space dimensions.} As shown in \cite{szwabowicz:pure} and \cite{saem:in-plane}, respectively, neither of these subclasses is empty.

\subsection{Pure Measure of Bending}\label{sec:def}
We say that a deformation measure $\pure(\y)$, of whatever nature (scalar or tensor), is a \emph{pure} measure of bending if it obeys the following invariance condition,
\begin{equation}
	\label{eq:pure_definition}
	\pure(\y^\ast)=\pure(\y),
\end{equation}
for all deformations $\y^\ast=\y'\circ\y$ such that $\y'$ is  bending-neutral.\footnote{A different definition for a \emph{natural} measure of bending was proposed in \cite{man:natural} in the case of infinitesimal incremental deformations of $\surface$.}

Since bending is intuitively associated with curvature, to flesh out this definition we first  need determine how $\normal^\ast$, the outer unit normal to $\surface^\ast$, is related to $\normal$ when $\y'$ is  a bending-neutral deformation (and then also how $\nabla\normal^\ast$ is related to $\nabla\normal$). To this end, we first recall that 
\begin{equation}
	\label{eq:normal_formula}
	\normal(\x)=\frac{\nay\e_1\times\nay\e_2}{|\nay\e_1\times\nay\e_2|}=\frac{\co\nay\e_3}{|\co\nay\e_3|},
\end{equation}
where $\co(\cdot)$ denotes the cofactor of a tensor (according to the definition  in \cite[$\S$\,2.11]{gurtin:mechanics}).
Since for any two tensors, $\Lt$ and $\M$, $\co(\Lt\M)=\co(\Lt)\co(\M)$, by \eqref{eq:chain_rule} and \eqref{eq:normal_formula}, we have that\footnote{Use is also made here of the fact that  $\co(\R)=\R$, for any rotation $\R\in\orth$.}
\begin{equation}
	\label{eq:normal_ast_formula}
	\normal^\ast=\frac{\co\naya\e_3}{|\co\naya\e_3|}=\frac{\co(\nabla\y')\normal}{|\co(\nabla\y')\normal|}=\R'\frac{\co(\U')\normal}{|\co(\U')\normal|}.
\end{equation}
Now, if $\y'$ is a bending-neutral deformation, since by \eqref{eq:U'} $\co(\U')=\lambda_1'\lambda_2'\normal\otimes\normal=(\det\U')\normal\otimes\normal$ and $\Rnu\normal=\normal$, we learn from \eqref{eq:normal_ast_formula} and \eqref{eq:bending-neutrality_definition} that 
\begin{equation}
	\label{eq:normal_ast}
	\normal^\ast=\R_0\normal.
\end{equation}
Moreover, since $\nabla\R_0=\zero$, we see that
\begin{equation}
	\label{eq:nabla_normal_ast}
	\nabla\normal^\ast=\R_0\nabla\normal.
\end{equation}

Thus, if we set
\begin{equation}\label{eq:A_definition}
	\A(\y):=\nan\trans\nan,
\end{equation}
we readily see from \eqref{eq:nabla_normal_ast} that this is a  pure (tensorial) measure of bending, as 
\begin{equation}
	\label{eq:A_ast}
\A(\y^\ast)=\nana\trans\nana=\nan\trans\nan=\A(\y).
\end{equation}
In the rest of the section, we shall focus on this measure. As defined in \eqref{eq:A_definition}, $\A$ is a linear mapping of the plane hosting the reference configuration $S$ into itself, precisely as $\C$ in \eqref{eq:representation_C}. Like the latter tensor, $\A$ is both symmetric and positive definite. According ot our definition, its scalar  invariants  are pure   measures of bending as well.

It is perhaps worth noting that the tensor $\mathbf{Z}(\y):=\nan\nan\trans$, similar in structure to $\B$, which maps the plane tangent to $\surface$ into itself, is \emph{not} a pure measure of bending, although it has the same invariants as $\A$: $\mathbf{Z}$ transforms  as follows, $\mathbf{Z}(\y^\ast)=\R_0\mathbf{Z}(\y)\R_0\trans$.

We denote by $\nablas\normal$ the \emph{curvature} tensor of $\surface$, for which we adopt  a sign convention opposite to the one customary in differential geometry.\footnote{Here we follow \cite{budiansky:notes} and \cite{sanders:nonlinear} in the desire to represent as $\nablas\normal=\frac{1}{R}\proj$ the curvature tensor of a \emph{sphere} of radius $R$.} $\nablas\normal$ is a symmetric tensor mapping the tangent plane to $\surface$ into itself; in its eigenframe, it is represented as
\begin{equation}
	\label{eq:representation_curvature}
	\nablas\normal=\kappa_1\bm{n}_1\otimes\bm{n}_1+\kappa_2\bm{n}_2\otimes\bm{n}_2,
\end{equation}
where $\bm{n}_1$, $\bm{n}_2$ are the principal directions of curvature of $\surface$ and $\kappa_1$, $\kappa_2$ are the corresponding principal curvatures.

By the chain rule,
\begin{equation}
	\label{eq:chain_rule_normals}
	\nabla\normal(\x)=\nablas\normal(\y(\x))\nabla\y(\x),
\end{equation}
and so it follows from \eqref{eq:A_definition} that $\A$ can also be written as 
\begin{equation}
	\label{eq:A_equivalent}
	\A=(\nabla\y)\trans\curvature^2(\nabla\y).
\end{equation}
Here and in the following, for brevity, we drop the argument from $\A(\y)$ and denote $\A^\ast:=\A(\y^\ast)$.

An interesting interpretation for $\A$ and its invariance under bending-neutral deformations can be given starting from \eqref{eq:A_equivalent}. This can indeed be seen as the referential representation for the \emph{third} fundamental form of $\surface$ (see, for example \cite[p.\,205]{grinfeld:introduction}). Given two arbitrary vectors $\vv_1$, $\vv_2$ on $S$,
\begin{equation}
	\label{eq:third_fundamental_form}
	\vv_1\cdot\A\vv_2=\vv_1'\cdot\curvature^2\vv_2'=\vv_1\cdot\A^\ast\vv_2=\vv_1^\ast\cdot(\nablas\normal^\ast)^2\vv_2^\ast,
\end{equation}	 
where $\vv_i'=\nay\vv_i$ are vectors tangent to $\surface$, $\vv_i^\ast=\naya\vv_i$ are vectors tangent to $\surface^\ast$, and $\nablas\normal^\ast$ is the curvature tensor of $\surface^\ast$. The identity \eqref{eq:third_fundamental_form} simply says that the third fundamental form is \emph{invariant} under a bending-neutral deformation.

As pointed out in \cite{acharya:nonlinear}, the tensorial deformation measure for shells considered in the works of Koiter~\cite{koiter:consistent}, Sanders~\cite{sanders:nonlinear}, and Budiansky~\cite{budiansky:notes} can be given a coordinate-free expression, which in the case of plates reads simply as
\begin{equation}
	\label{eq:K_definition}
	\mathbf{K}:=\nay\trans\curvature\nay.
\end{equation}
This appears to be the referential representation of the \emph{second} fundamental form of $\surface$. In contrast with $\A$, $\mathbf{K}$ is not a pure measure of bending. 

The linear invariant of $\A$, $\tr\A$ (which is quadratic in the principal curvatures of $\surface$), has already featured in the literature; it emerged as a strain energy density from  a discretized model for membranes and plates \cite{seung:defects} and, more recently, also from a dimension reduction to plates performed in \cite{vitral:energies}. In the latter, the reduced strain energy density also involves the scalar invariants of the tensor $\bm{\Pi}:=\frac12(\V\curvature+\curvature\V)$, where $\V$ is the  stretching tensor of $\y$ in the present configuration, defined as the positive definite root of $\V^2=\B$. It is a simple exercise to show by example that neither $\bm{\Pi}$ nor its invariants are pure measures of bending according to the definition proposed in this paper.

Finally, the shell model advanced in \cite{knoche:buckling} employs strains that reduce to the eigenvalues of $\A$ in the case of plates, under the further hypothesis that $\V$ and $\nablas\normal$ commute, a hypothesis implicit in the symmetric class of deformations adopted in \cite{knoche:buckling}.\footnote{This hypothesis, however, fails to make $\bm{\Pi}$ a pure measure of bending for plates.}   

\subsection{Impure Measures of Bending}\label{sec:impure}
It is interesting to link the invariants of $\A$ to other invariant, but impure measures of bending, namely, the mean curvature $H$ and the Gaussian curvature $K$ of $\surface$, defined as
\begin{equation}
	\label{eq:curvature_invariants}
	H:=\frac12\tr\curvature\quad\text{and}\quad K:=\det\curvature.
\end{equation}

We begin by computing $\tr\A$. It 
readily follows from \eqref{eq:A_equivalent} that 
\begin{equation}
	\label{eq:trA}
	\tr\A=\tr(\B\curvature^2)=2H\tr(\B\nablas\normal)-K\tr\B,
\end{equation}
where the last equality follows from an identity proven in \cite{ozenda:kirchhoff} (see their equation (14)).

To obtain $\det\A$, we start from $\tr\A^2$, which by \eqref{eq:A_equivalent}
also reads as
\begin{equation}\label{eq:trA^2}
	\tr\A^2=\curvature^2\B\cdot\B\curvature^2,	
\end{equation}
where $\cdot$ denotes the inner product of tensors.\footnote{For generic  tensors $\Lt$ and $\M$, $\Lt\cdot\M:=\tr(\Lt\trans\M)$.} Letting $\B$ be represented as
\begin{align}
	\B&=B_{11}\bm{n}_1\otimes\bm{n}_1+B_{12}(\bm{n}_1\otimes\bm{n}_2+\bm{n}_2\otimes\bm{n}_1)+B_{22}\bm{n}_2\otimes\bm{n}_2\label{eq:B_representation}
\end{align}
in the eigenframe of $\nablas\normal$ in \eqref{eq:representation_curvature},
we easily obtain from \eqref{eq:trA^2} that
\begin{equation}
	\label{eq:trA^2_final}
	\tr\A^2=\kappa_1^4B_{11}^2+\kappa_2^4B_{22}^2+2\kappa_1^2\kappa_2^2B_{12}^2=(\tr\A)^2-2K^2\det\B,
\end{equation}
which by the Cayley-Hamilton theorem applied to $\A$ implies that 
\begin{equation}
	\label{eq:det_A}
	\det\A=K^2\det\B=K^2\det\C.
\end{equation}
Thus, equations \eqref{eq:trA} and \eqref{eq:det_A} show how pure scalar  measures of bending may result from combining measures of stretching and impure measures of bending.

In the special case of an isometry, for which $\C=\I-\e_3\otimes\e_3$ and $\B=\proj$, 
\begin{equation}
	\label{eq:isometry}
	\tr\A=\kappa_1^2+\kappa_2^2=2C\quad\text{and}\quad\det\A=(\kappa_1\kappa_2)^2=K^2,
\end{equation} 
where $C$ is the Casorati total curvature \cite{casorati:mesure}.

\subsection{Pure Rod-Bending Measure}\label{sec:rod}
\begin{figure}[h]
	\begin{center}
		\includegraphics[width=.4\linewidth]{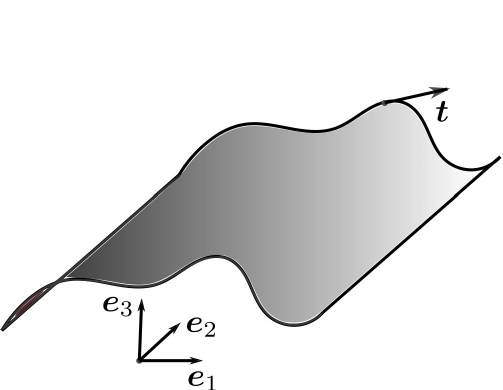}
	\end{center}
	\caption{Cylindrical surface produced by the deformation $\y$ described in \eqref{eq:y_cylinder}.}
	\label{fig:cylinder}
\end{figure}

To connect $\A$ to Antman's measure of pure bending for rods \cite{antman:general}, we consider a class of deformations of $S$ that produce a cylindrical surface, as shown in Fig.~\ref{fig:cylinder}.
More precisely, we take
\begin{equation}
	\label{eq:y_cylinder}
	\y(x_1,x_2)=y_1(x_1)\e_1+x_2\e_2+y_3(x_1)\e_3,
\end{equation}
where $y_1$, $y_3$ are scalar functions of class $C^2$. From \eqref{eq:y_cylinder} we easily arrive at
\begin{equation}
	\label{eq:nabla_y_cylinder}
	\nabla\y=y_1'\e_1\otimes\e_1+\e_2\otimes\e_2+y_3'\e_3\otimes\e_1,
\end{equation}
where a prime $'$ now denotes differentiation with respect to $x_1$. Thus, by applying \eqref{eq:normal_formula}, we see that 
\begin{equation}
	\label{eq:normal_cylinder}
	\normal=\cos\vt\e_3-\sin\vt\e_1\quad\text{with}\quad\cos\vt=\frac{y_1'}{\sqrt{y_1'^{2}+y_3'^{2}}}\quad\text{and}\quad\sin\vt=\frac{y_3'}{\sqrt{y_1'^{2}+y_3'^{2}}}.
\end{equation}
With this choice of the angle $\vt$, the unit tangent $\tangent$ to the curve in space that sweeps $\surface$ along the generatrix $\e_2$ is represented as
\begin{equation}
	\label{eq:tangent}
	\tangent=\frac{\nay\e_1}{|\nay\e_1|}=\cos\vt\e_1+\sin\vt\e_3,
\end{equation}
and it follows from \eqref{eq:normal_cylinder} that
\begin{equation}
	\label{eq:A_cylinder}
	\A=\vt'^2\e_1\otimes\e_1.
\end{equation}
The only non-vanishing invariant of $\A$ is $\tr\A=\vt'^2$, which, to within an elastic modulus, is precisely Antamn's bending energy for an extensible elastic rod.

\nigh{
\subsection{An Illustration: Axisymmetric Deformations}
To illustrate the theory proposed in this paper, we consider now the general class of axisymmetric deformations of a \emph{disk} of radius $R$. The geometric setting is shown in Fig.~\ref{fig:nightcap}, where $S$ is the undeformed disk and $\surface$ is an axisymmetric \emph{nightcap} generated by the deformation $\y$ parametrized as
\begin{equation}
	\label{eq:y_axisymmetric}
	\y(s,\varphi)=r(s)\e_r+z(s)\e_3,
\end{equation}
where $(s,\vp)$, with $0\leqq s\leqq R$ and $0\leqq\vp\leqq2\pi$, are polar coordinates in the plane $(x_1,x_2)$, so that $x_1=s\cos\vp$ and $x_2=s\sin\vp$, $\e_r$ is the radial unit vector and $\e_\vp$ the azimuthal unit vector, $r(s)$ and $z(s)$ are smooth functions representing radial and vertical components of the image under  $\y$ of the generic point $\x=s\e_\vp$ of $S$.
\begin{figure}[h]
	\begin{center}
		\includegraphics[width=.35\linewidth]{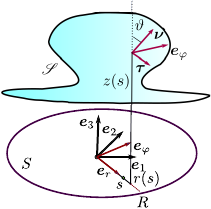}
	\end{center}
	\caption{\nigh{Sketch of an axisymmetric \emph{nightcap} $\surface$ obtained by deforming a flat disk $S$ of radius $R$ via the mapping $\y$ described by \eqref{eq:y_axisymmetric}. The unit vector $\normal$ is the outward normal to $\surface$, while $\bm{\tau}$ is the tangent unit vector along the meridians of $\surface$, so that $(\bm{\tau},\e_\vp,\normal)$ is a mobile frame oriented like the fixed frame $\framec$. The angle $\vt$, defined in \eqref{eq:theta_definition}, designates the relative inclination of the normals $\normal$ and $\e_3$, to $\surface$ and $S$, respectively.}}
	\label{fig:nightcap}
\end{figure}

Let $\x(t)=s(t)\e_r(\vp(t))$ be a generic curve on $S$ parameterized in $t$. Denoting by a superimposed dot differentiation with respect to $t$, we readily obtain that
\begin{equation}
	\label{eq:curve_tangents}
	\dot{\x}=\dot{s}\e_r+s\dot{\vp}\e_\vp\quad\text{and}\quad\dot{\y}=\dot{s}\left(r'\e_r+z'\e_3\right)+r\dot{\vp}\e_\vp,
\end{equation}
where a prime $'$ denotes differentiation with respect to $s$ and $\dot{\y}$ is the tangent to the curve $\y(t)$ resulting from composing $\y$ with $\x(t)$. The deformation gradient $\nabla\y$ at a point $\x_0$ must satisfy the identity $\dot{\y}=\nay\dot{\x}$ for every curve $\x(t)$ through $\x_0$. The following formula is then an easy consequence of \eqref{eq:curve_tangents},
\begin{equation}
	\label{eq:nabla_y_axisymmetric}
	\nabla\y=r'\e_r\otimes\e_r+\frac{r}{s}\e_\vp\otimes\e_\vp+z'\e_3\otimes\e_r.
\end{equation}
It follow from \eqref{eq:nabla_y_axisymmetric} that 
\begin{equation}
	\label{eq:C_axysimmetric}
	\C=\nay\trans\nay=\lambda_1^2\e_r\otimes\e_r+\lambda_2^2\e_3\otimes\e_3,
\end{equation}
where
\begin{equation}
	\label{eq:stretches_axisymmetric}
	\lambda_1:=\sqrt{r'^2+z'^2}\quad\text{and}\quad\lambda_2:=\frac{r}{s} 
\end{equation}
are the principal stretches of $\U=\sqrt{\C}$.

If the top of the nightcap in Fig.~3 is regular, that is, with a well defined tangent plane, symmetry requires the latter to be parallel to the $(x_1,x_2)$ plane, and so $\nabla\y$ must be the identity in that plane. From (35), this implies that $r'(0)=\lim_{s\to0}r(s)/s=1$ and $z'(0)=0$, meaning that $\lambda_1(0)=\lambda_2(0)=1$.

A simple computation yields
\begin{equation}
	\label{eq:co_nabla_y_axisymmetric}
	\co(\nabla\y)=\frac{r}{s}\left(r'\e_3\otimes\e_3-z'\e_r\otimes\e_3\right),
\end{equation}    
which by \eqref{eq:normal_formula} leads us to
\begin{equation}
	\label{eq:normal_axisymmetric}
	\normal=\frac{1}{\lambda_1}\left(r'\e_3-z'\e_r\right).
\end{equation}
Defining by $\bm{\tau}:=\e_\vp\times\normal$ a tangent unit vector to $\surface$ oriented along its meridians, so that the mobile frame $(\bm{\tau},\e_\vp,\normal)$ is oriented like $\framec$, we can write
\begin{equation}
	\label{eq:tangent_axisymmetric}
	\bm{\tau}=\frac{1}{\lambda_1}\left(r'\e_r+z'\e_3\right),
\end{equation}
so that \eqref{eq:nabla_y_axisymmetric} becomes
\begin{equation}
	\label{eq:nabla_y_axisymmetric_concise}
	\nabla\y=\lambda_1\bm{\tau}\otimes\e_r+\lambda_2\e_\vp\otimes\e_\vp.
\end{equation}
From this latter we easily compute
\begin{equation}
	\label{eq:B_axisymmetric}
	\B=\nay\nay\trans=\lambda_1^2\bm{\tau}\otimes\bm{\tau}+\lambda_2^2\e_\vp\otimes\e_\vp
\end{equation}
and
\begin{equation}
	\label{eq:nabla_y_inverse_axisymmetric}
	\nay^{-1}=\frac{1}{\lambda_1}\e_r\otimes\bm{\tau}+\frac{1}{\lambda_2}\e_\vp\otimes\e_\vp.
\end{equation} 

By introducing the angle
\begin{equation}
	\label{eq:theta_definition}
	\vt(s):=-\arctan\frac{z'}{r'},
\end{equation}
we give a more compact representation of $\normal$ and $\bm{\tau}$,
\begin{equation}
	\label{eq:nu_and_tau}
	\normal=\cos\vt\e_3+\sin\vt\e_r,\quad\bm{\tau}=\cos\vt\e_r-\sin\vt\e_3.
\end{equation}
As a consequence of \eqref{eq:nu_and_tau} and \eqref{eq:nabla_y_axisymmetric_concise}, we can write
\begin{equation}
\label{eq:nabla_normal_axisymmetric}
\nabla\normal=\vt'\bm{\tau}\otimes\e_r+\frac{1}{s}\sin\vt\e_\vp\otimes\e_\vp,
\end{equation}
from which it follows that the pure measure of bending $\A$ in \eqref{eq:A_definition} acquires here the explicit representation
\begin{equation}
	\label{eq:A_axisymmetric}
	\A=\vt'^2\e_r\otimes\e_r+\frac{1}{s^2}\sin^2\vt\e_\vp\otimes
	\e_\vp,
\end{equation}
so that a quadratic pure bending invariant would be
\begin{equation}
	\label{eq:trA_axisymmetric}
	\tr\A=\vt'^2+\frac{1}{s^2}\sin^2\vt.
\end{equation}

For a comparison, we also compute the quadratic invariants of the curvature tensor $\nablas\normal$. By \eqref{eq:chain_rule_normals}, \eqref{eq:nabla_y_inverse_axisymmetric}, and \eqref{eq:nabla_normal_axisymmetric}, we obtain that 
\begin{equation}
	\label{eq:curvature_tensor_axisymmetric}
	\nablas\normal=\nan\nay^{-1}=\frac{1}{\lambda_1}\vt'\bm{\tau}\otimes\bm{\tau}+\frac{1}{\lambda_2s}\sin\vt\e_\vp\otimes\e_\vp,
\end{equation}
which, compared with  \eqref{eq:representation_curvature}, yields
\begin{equation}
	\label{eq:principal_curvatures_axisymmetric}
	\kappa_1=\frac{\vt'}{\sqrt{r'^2+z'^2}}\quad\text{and}\quad\kappa_2=\frac{\sin\vt}{r}.
\end{equation}
The quadratic impure bending measures would then read as
\begin{equation}
	\label{eq:H_K_axisymmetric}
	H^2=\frac14\left(\frac{\vt'}{\sqrt{r'^2+z'^2}}+\frac{\sin\vt}{r}\right)^2\quad\text{and}\quad K=\frac{\vt'\sin\vt}{r\sqrt{r'^2+z'^2}}.
\end{equation}
By combining together \eqref{eq:B_axisymmetric}, \eqref{eq:A_axisymmetric}, and the expression for $K$ in \eqref{eq:H_K_axisymmetric}, one easily sees that identity \eqref{eq:det_A} is indeed satisfied.

A simple direct theory of isotropic soft plates developed according to the principles purported in this paper would be based on the following strain energy density (per unit reference area),
\begin{eqnarray}
	\label{eq:strain_energy_density_axisymmetric}
	W&:=&\frac12\alpha\tr\A+\frac12\beta[\tr\U^2+2(1-\tr\U)]\nonumber\\
	&=&\frac12\alpha\left(\vt'^2+\frac{1}{s^2}\sin^2\vt\right)+\frac12\beta[(\lambda_1-1)^2+(\lambda_2-1)^2],
\end{eqnarray}
where $\alpha$ and $\beta$ are positive elastic moduli. A similar (but not identical) expression for $W$ was proposed for axisymmetric shells in \cite{cohen:axisymmetric} (see also \cite[p.\,642]{villaggio:mathematical}); it was amenable to a formal Hamiltonian treatment. The strain energy functional resulting from $W$ is then
\begin{equation}
	\label{eq:strain_energy_functional_axisymmetric}
	F[r,z]:=\int_S W\dd A=\pi\int_0^R\left\{\alpha\left(\vt'^2+\frac{1}{s^2}\sin^2\vt\right)+\beta[(\lambda_1-1)^2+(\lambda_2-1)^2]\right\}s\dd s,
\end{equation}
subject to appropriate boundary conditions on the edge of $S$.

By contrast, a quadratic theory building on the impure measures in \eqref{eq:H_K_axisymmetric}, would be based on an Helfrich-type functional,
\begin{eqnarray}
	\label{eq:Helfrich_energy_functional_axisymmetric}
	F_\mathrm{H}[r,z]&:=&\int_{\surface}(\alpha H^2+\beta K)\dd a\nonumber\\
	&=&\int_S\lambda_1\lambda_2\left[\frac14\alpha\left(\frac{\vt'}{\lambda_1}+\frac{\sin\vt}{r}\right)^2+\beta\frac{\vt'}{\lambda_1}\frac{\sin\vt}{r}\right]\dd A\nonumber\\
	&=&\pi\int_0^R\left[\frac12\alpha\left(\frac{r}{\sqrt{r'^2+z'^2}}\vt'^2+\frac{\sqrt{r'^2+z'^2}}{r}\sin^2\vt\right)+(2\beta+\alpha)\vt'\sin\vt\right]\dd s,
\end{eqnarray}
where the area element $\dd a$ on the present shape $\surface$ has been related to the area element $\dd A=s\dd s\dd\vp$ on the reference shape $S$ through the equation $\dd a=\lambda_1\lambda_2\dd A$.

Two features deserve notice in \eqref{eq:Helfrich_energy_functional_axisymmetric}. First, as appropriate for a liquid membrane, $F_\mathrm{H}$ is formulated as an integral on the present configuration $\surface$. Second, the Gaussian curvature $K$ reveals itself as a \emph{null Lagrangian}, as its contribution to the total strain energy depends only on the boundary value of $\vt$,
\begin{equation}
	\label{eq:null_Lagrangian}
	\int_{\surface}K\dd a=2\pi[1-\cos\vt(R)],
\end{equation}
thus becoming an effective \emph{edge} energy.\footnote{The interplay between $F_\mathrm{H}$ and edge strain energies has been explored in a series of papers, some also recent \cite{boal:topology,capovilla:lipid,tu:compatibility,tu:geometry,tu:recent,zhou:integral,palmer:minimal,palmer:minimizing,palmer:euler}.}
}

\section{Bending-Neutrality Compatibility}\label{sec:compatibility}
For $\y'$ to be a bending-neutral deformation, neither $\U'$ nor $\Rnu$ can be arbitrary. They must be subject to a compatibility condition, which we shall discuss in some detail here along with its solutions. The special case where $\U'=\proj$ has been considered and fully solved in \cite{saem:in-plane}.

Compatibility arises from requiring that the curvature tensor $\nablas\normal^\ast$ of $\surface^\ast$ be symmetric, as it should (see also \cite{saem:in-plane}). By writing \eqref{eq:chain_rule_normals} for $\normal^\ast$ (and $\y^\ast$) instead of $\normal$ (and $\y$), we have that 
\begin{equation}
	\label{eq:first_remark}
	\nabla\normal^\ast=(\nablas\normal^\ast)(\nabla\y^\ast).
\end{equation}
On the other hand, by  \eqref{eq:normal_ast} and \eqref{eq:nabla_normal_ast}, we can also write
\begin{equation}
	\label{eq:second_remark}
	\nabla\normal^\ast=\R_0\nabla\normal=\R_0(\nablas\normal)\nay,
\end{equation}
which, combined with \eqref{eq:first_remark}, leads us to
\begin{equation}
	\label{eq:conclusion_remark}
	\nablas\normal^\ast=\R_0\curvature(\U')^{-1}\Rnu\trans\R_0\trans,
\end{equation}
once use is also made of \eqref{eq:chain_rule}, \eqref{eq:nabla_y'}, and \eqref{eq:bending-neutrality_definition}.
It follows from \eqref{eq:conclusion_remark} that $\nablas\normal^\ast$ is symmetric if and only if $\curvature(\U')^{-1}\Rnu\trans$ is so. This latter condition will be written in the following equivalent way,
\begin{equation}
	\label{eq:compatibility_condition}
	\nablas\normal=\Rnu\Upm\curvature\Rnu\U'.
\end{equation}
Tensors on both sides of \eqref{eq:compatibility_condition} act on the local tangent plane of $\surface$. It is clear that every solution $\U'$ of \eqref{eq:compatibility_condition} is defined to within a multiplicative surface dilation, $\lambda'\proj$, with arbitrary $\lambda'>0$.

Two special classes of solutions of \eqref{eq:compatibility_condition} deserve notice. They are better illustrated by use of the classical Euler-Rodriguez formula to represent $\orthnu$ (see also \cite{palais:disorienting} for an updated account),
\begin{equation}
	\label{eq:E_R_formula}
	\R(\alpha):=\I+\sin\alpha\W(\normal)-(1-\cos\alpha)\proj,\quad\alpha\in]-\pi,\pi],
\end{equation}
where $\W(\normal)$ is the skew-symmetric tensor associated with $\normal$, so that $\W(\normal)\vv=\normal\times\vv$, for all vectors $\vv$.

First, if either $\Rnu=\R(0)=\I$ or $\Rnu=\R(\pi)=-\I+2\normal\otimes\normal$, the latter acting as $-\I$ on the plane tangent to $\surface$, then \eqref{eq:compatibility_condition} reduces to requiring that $\U'$ commute with $\nablas\normal$, which just amounts to say that $\nablas\normal$ and $\U'$ must have the same eigenframe (see also \cite{szwabowicz:pure}).

Second, for $\U'=\lambda'\proj$, with $\lambda'>0$, \eqref{eq:compatibility_condition} becomes
\begin{equation}
	\label{eq:compatibility_second}
	\nablas\normal=\Rnu\curvature\Rnu.
\end{equation}
Taking the trace of both sides of \eqref{eq:compatibility_second}, and using both \eqref{eq:representation_curvature} and \eqref{eq:E_R_formula}, we arrive at
\begin{equation}
	\label{eq:trace_formula}
	2H(1-\cos2\alpha)=0,
\end{equation}
which for $2H=\tr\curvature\neq0$ reduces to either $\alpha=0$ or $\alpha=\pi$, delivering again the first special solution of \eqref{eq:compatibility_condition} considered above. For $H=0$, which is the case of \emph{minimal} surfaces, \eqref{eq:trace_formula} is identically satisfied and direct inspection of \eqref{eq:compatibility_second} with the aid of \eqref{eq:E_R_formula} shows that it is solved by all $\Rnu\in\orthnu$. For $\lambda'=1$, this result was already proved in \cite{saem:in-plane}.\footnote{As appropriately remarked in \cite{saem:in-plane}, the same result had also been proved, albeit with different methods, in the differential geometry literature (see \cite{abe:isometric,hoffman:Gauss,eschenburg:compatibility}), but it was unknown to plates and shells practitioners.}

To find the general solution of \eqref{eq:compatibility_condition}, we represent $\U'$ in \eqref{eq:U'} in the eigenframe of $\nablas\normal$ in \eqref{eq:representation_curvature}, by writing
\begin{equation}
	\label{eq:U'_alternative}
	\U'=\R(\chi)\U_0'\R(-\chi),
\end{equation}
where $\R(-\chi)=\R(\chi)\trans$,
\begin{equation}
	\label{eq:U'_0}
	\U_0':=\lambda_1'\bm{n}_1\otimes\bm{n}_1+\lambda_2'\bm{n}_2\otimes\bm{n}_2,
\end{equation}
and $\chi$ is the angle that $\uv_1'$ makes with $\bm{n}_1$ (the same as the angle that $\uv_2'$ makes with $\bm{n}_2$). By use of these representation formulae and \eqref{eq:E_R_formula}, we finally give \eqref{eq:compatibility_condition} the following expanded form,
\begin{equation}
	\label{eq:compatibility_parameters}
	\nablas\normal=\R(\alpha+\chi)(\U_0')^{-1}\R(-\chi)\curvature\R(\alpha+\chi)\U_0'\R(-\chi),
\end{equation}
which, for given $\kappa_1$, $\kappa_2$, is meant to be an equation for $\lambda_2'/\lambda_1'$, $\alpha$, and $\chi$.

Actually, it is not difficult to show that \eqref{eq:compatibility_parameters} amounts to four scalar equations, whose solution is
\begin{equation}
	\label{eq:compatibility_solution}
	\frac{\lambda_2'}{\lambda_1'}=1-\frac{(\kappa_1+\kappa_2)\sin\alpha}{\kappa_2\sin\alpha+(\kappa_1-\kappa_2)\cos\chi\sin(\alpha+\chi)}.
\end{equation}
Requiring the ratio $\lambda_2'/\lambda_1'$ as delivered by \eqref{eq:compatibility_solution} to be positive, with little labour we arrive at the inequalities
\begin{equation}
	\label{eq:compatibility_inequalities}
	\alpha_\mathrm{m}<\alpha<\alpha_\mathrm{M},
\end{equation}
where $\alpha_\mathrm{m}(\chi):=\min\{\alpha_1,\alpha_2\}$ and $\alpha_\mathrm{M}(\chi):=\max\{\alpha_1,\alpha_2\}$, with
\begin{equation}
	\label{eq:alpha_s}
	\alpha_1:=\arctan\left(\frac{(\kappa_2-\kappa_1)\cos\chi\sin\chi}{\kappa_1\sin^2\chi+\kappa_2\cos^2\chi}\right)\quad\text{and}\quad\alpha_2:=\arctan\left(\frac{(\kappa_2-\kappa_1)\cos\chi\sin\chi}{\kappa_1\cos^2\chi+\kappa_2\sin^2\chi}\right).
\end{equation}
Thus, for given $\chi\in]-\pi,\pi[$ and $\alpha\in]\alpha_\mathrm{m},\alpha_\mathrm{M}[$, $\U'$ is delivered by \eqref{eq:U'_alternative} to within a surface dilation $\lambda'\proj$; in general, there is a whole three-dimensional set of solutions to \eqref{eq:compatibility_condition}, parameterized by $(\chi,\alpha,\lambda')$. A singular case arises at umbilical points of $\surface$, where $\kappa_1=\kappa_2$. There, \eqref{eq:compatibility_solution} would be incompatible for any $\alpha\notin\{0,\pi\}$ since it requires $\lambda_2'/\lambda_1'=-1$, whereas, by \eqref{eq:alpha_s}, $\alpha_\mathrm{m}=\alpha_\mathrm{M}=0$, so that  the standard solution $\U'=\lambda'\proj$, $\Rnu=\I$ is recovered by continuity.

A particular two-dimensional family of solutions deserves attention. It is obtained from \eqref{eq:compatibility_solution} by setting $\chi=0$, under the assumption that $\alpha\notin\{0,\pi\}$ (an equivalent family would be obtained by setting $\chi=\frac\pi2$). If $\surface$ is a \emph{hyperbolic} surface, that is, if $K<0$, \eqref{eq:compatibility_solution} readily gives
\begin{equation}
	\label{eq:compatibility_hyperbolic}
	\frac{\lambda_2'}{\lambda_1'}=-\frac{\kappa_2}{\kappa_1},
\end{equation}
while $\alpha$ is arbitrary. What makes this solution special is the curvature tensor of the surface $\surface^\ast$ delivered by $\y'$. Letting $\R_0=\I$, with no prejudice to generality, we obtain from \eqref{eq:conclusion_remark} that 
\begin{equation}
	\label{eq:curvature_minimal_surface}
	\nablas\normal^\ast=\frac{\kappa_1}{\lambda_1'}\bm{n}_1\otimes\R(\alpha)\bm{n}_1+\frac{\kappa_2}{\lambda_2'}\bm{n}_2\otimes\R(\alpha)\bm{n}_2,
\end{equation} 
whence, also by \eqref{eq:compatibility_hyperbolic}, it follows that 
\begin{subequations}
\begin{align}
2H^\ast&:=\tr\curvaturast=\cos\alpha\left(\frac{\kappa_1}{\lambda_1'}+\frac{\kappa_2}{\lambda_2'}\right)=0,	\label{eq:mean_curvature}\\
K^\ast&:=\det\curvaturast=-\left(\frac{\kappa_1}{\lambda_1'}\right)^2=-\left(\frac{\kappa_2}{\lambda_2'}\right)^2.\label{eq:gaussian_curvature}
\end{align}
\end{subequations}
In particular, \eqref{eq:mean_curvature} implies that $\surface^\ast$ is a minimal surface.

Clearly, this conclusion relies on the very existence of $\surface^\ast$, which would be guaranteed, at least locally, if the surface tensor field defined on $\surface$ as
\begin{equation}
\label{eq:S_definition}
\mathbf{S}:=\lambda'\R(\alpha)\U'
\end{equation} 
were a surface \emph{gradient}. This is an integrability requirement that poses restrictions on both $\lambda'$ and $\alpha$, which remain the only unknown scalar surface fields, once \eqref{eq:compatibility_hyperbolic} is enforced. The study of these restrictions exceeds the scope of this paper and will be the subject of a future one, as will also be the integrability of $\mathbf{S}$ for the general class of incremental deformations that obey \eqref{eq:compatibility_solution}.

We have considered two distinct conditions that must be met for $\y'$ to be a bending-neutral deformation, one compatibility condition and one integrability condition. Whenever $\surface$ is such that either one or the other condition is not met, any incremental  $\y'$ will necessarily bring additional bending.

\section{Conclusions}\label{sec:conclusions}
The primary object of this paper is answering a question posed in \cite{wood:contrasting}:``How do we wish to define a measure of bending, and by extension a bending energy and the concept of pure stretching deformation, for a thin structure''?

We (the patient reader and I) caracolled  our way toward a possible answer through the notion of bending-neutral deformation, an incremental deformation that does not affect bending, but it is more than a pure stretching as 
it may include normal twisting. 

Quite naturally, a pure measure of bending was then introduced  as a deformation measure unaffected by bending-neutral deformations. An explicit, tensorial example of such a measure was given and contrasted against other measures known for plates.

Common wisdom has it that plates (and shells, for that matter) are rather unresponsive to twists exciting local drill rotations (see, for example, \cite{szwabowicz:pure}). This may well be the case for conventional, \emph{hard} plates, but it is likely not to be so for \emph{soft}, polymeric plates, especially those with an activable internal structure, such as nematic elastomers (described in the landmark textbook \cite{warner:liquid}).

These latter material surfaces are \emph{anisotropic}, as the nematic director field $\bm{n}$ on $\surface$ breaks the rotational symmetry about the normal $\normal$. It might be interesting to find anisotropic pure measure of bending fit for these systems; they might be suggestive of direct theories where stretching and bending energy contents are kept separate, oblivious to whether they can be derived by dimension reduction of established three-dimensional strain energies or not.\footnote{The role of bending energy in the theory for nematic elastomer plates has been highlighted in \cite{ozenda:blend}; applications are presented in \cite{singh:ribbon,singh:bending} and \cite{sonnet:model}.} 

Whether the answer presented in this paper  to  the opening question is satisfactory or not is not for me to tell. In the second part of the paper, my interest shifts toward the legitimacy of bending-neutral deformations. They have been characterized by the fulfillment of a necessary kinematic compatibility condition, which generates a three-dimensional set of solutions. Of course, this only represents the admissible bending-neutral deformations; which ones do actually exist, given a surface $\surface$ they are meant to transform, is decided, at least locally, by the fulfillment of an integrability condition. The study of this condition and its solutions will be the subject of a future paper. 

%

\end{document}